# Integrated Sensing and Communication in 3GPP: Evolution from 5G-Advanced to 6G

Xingqin Lin

NVIDIA

Email: xingqinl@nvidia.com

***Abstract*— Integrated sensing and communication (ISAC) extends mobile networks from information transfer toward perception of passive objects and environments. By exploiting propagation delay, Doppler, angle, and temporal variation, a mobile network can support detection and tracking while reusing licensed spectrum, infrastructure, and edge computing at network scale and under operator control. Unlike conventional radar, ISAC must additionally address multi-application access, heterogeneous sensing entities, uncertainty, privacy, trust, and integration with communication services. This article reviews the evolution of ISAC in 3GPP from 5G-Advanced to 6G. It explains Release-19 ISAC service requirements and channel models, the Release-20 sensing function architecture and reporting abstractions, and the 5G monostatic baseline for drone detection and tracking. It also examines 6G-native sensing, including passive object sensing, communication assistance, multiple sensing modes, and multi-source data integration. The article makes the evolution explicit through a comparison of sensing capabilities and KPIs, an assessment of reusable functions and deployment costs, and a quantitative UAV case study linking service targets to sensing configuration, reporting overhead, and delivery latency.**

## I. INTRODUCTION

Integrated sensing and communication (ISAC) uses propagation delay, Doppler, spatial phase, and temporal channel variations to observe objects and environments through mobile radio signals [1]. A sensing receiver may process reflected signals to detect an uncrewed aerial vehicle (UAV), estimate its state, and support tracking. Other potential targets include vehicles, pedestrians, railway obstacles, automated guided vehicles (AGVs), and industrial machinery. The target need not carry an active radio device.

ISAC must support sensing across multiple applications, radio nodes, and processing locations [2]. An external application may request observations that are generated at a base station, processed near the radio access network (RAN), fused across sites, and exposed through the operator network. The system must authorize the request, select suitable entities and resources, and choose whether to transport channel information, profiles, detected points, or object results. Timestamping, uncertainty, privacy, and access control are therefore part of the service design.

Passive-object sensing differs from cellular positioning of a participating device. A passive target may transmit nothing and is observed through reflections, blockage, or diffraction of network signals. Processing must distinguish its response from clutter and interference, including local transmitter leakage when monostatic transmission and reception overlap. Device-free observation also introduces privacy and trust concerns for targets without a network subscription.

Existing surveys establish the foundations of ISAC, including joint waveform and receiver design, sensing topologies, network architectures, and infrastructure reuse [1], [2]. Ghosh *et al.* discuss the broader 6G ISAC opportunity and associated radio and implementation challenges [3]. More recently (after this article was made public), Varshney reviews ISAC evolution in 3GPP across service, physical-layer, protocol, and architectural aspects [4]. Building on the literature, the distinctive contribution here is an explicit comparison of 5G-Advanced and 6G sensing objectives; an assessment of reusable functions, necessary extensions, and deployment costs; and a quantitative case study linking service requirements to sensing configuration, processing placement, reporting traffic, and delivery delay.

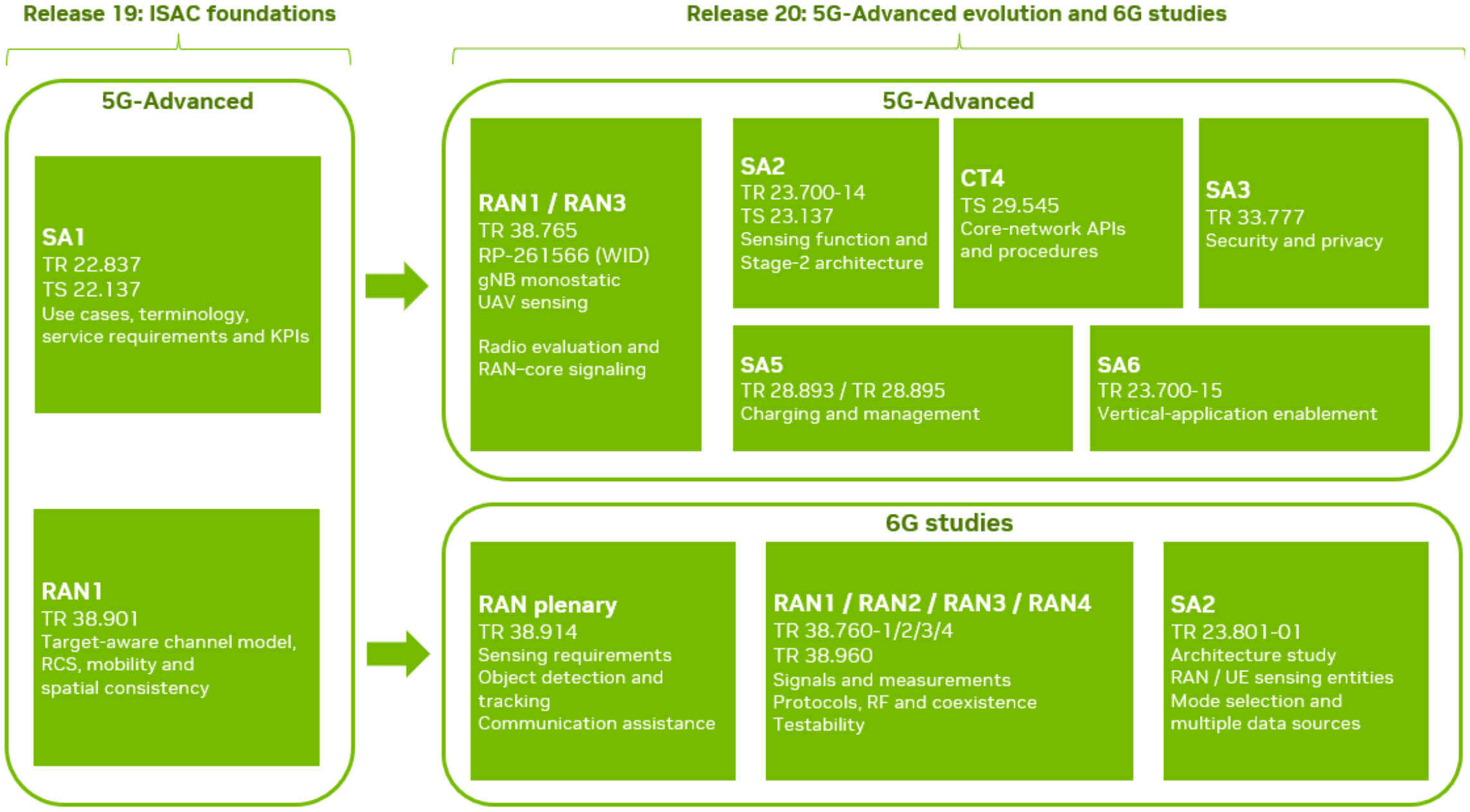

**Figure 1: Evolution of 3GPP ISAC from Release 19 foundations to the dual 5G-Advanced and 6G tracks in Release 20.**

The 3rd Generation Partnership Project (3GPP) initiated ISAC work within 5G-Advanced in Release 19, while Release 20 combines continued 5G-Advanced evolution with the first 6G studies, as illustrated in Figure 1. In Release 19, service and system aspects (SA) working group 1 studied ISAC use cases and established normative service requirements [5], [6]. In Release 20, TS 23.137 specifies the 5G system (5GS) sensing architecture and service procedures [7]. Related work addresses security and privacy [8], management and charging, vertical-application enablement [9], core-network application programming interfaces (APIs), and RAN-core signaling. On the RAN side, Release-19 work extended the channel model with sensing target, radar cross section (RCS), mobility, and spatial-consistency aspects [10]. The completed 5G new radio (NR) study TR 38.765 evaluates 5G node B (gNB) based monostatic UAV sensing using existing downlink waveforms and reference signals [11]. Its evaluation findings and architectural observations inform subsequent specification work. In parallel, Release-20 6G studies broaden the design space. TR 38.914 includes passive-object sensing and communication assistance among the 6G scenarios and requirements [12]. The SA2 architecture study TR 23.801-01 considers RAN and user equipment (UE) sensing entities, six sensing modes, and integration of non-3GPP sensing information [13]. The ongoing 6G radio study examines how physical-layer design, higher-layer procedures, radio frequency (RF) constraints, coexistence, and testability can support these broader objectives.

In this article, we first explain sensing services, topologies, performance metrics, and the targeted evolution from 5G-Advanced to 6G. We then examine the 5GS sensing architecture, processing abstractions, 6G extensions, and deployment costs. The radio discussion retains target-aware channel modeling and a tutorial processing chain, followed by an assessment of bandwidth and frequency choices. A UAV case study connects the service and implementation layers before we conclude with open challenges. The standardization discussion uses the cited document versions and reflects a snapshot as of 15 October 2026 [5]–[13]. Table 1 distinguishes specification provisions, completed studies, ongoing studies, and proposals.

## II. SENSING SERVICES AND REQUIREMENTS

### *A. Service Model and Use Cases*

The 3GPP sensing service model separates the physical radio observation from the information delivered to a consumer. A sensing signal is a transmission over the 3GPP radio interface that can be used to observe an object or environment. After interaction with the physical scene, a sensing receiver derives sensing data, such as channel estimates, delay or Doppler profiles, angular spectra, or detected points. Further processing generates a sensing result, such as object presence, position, velocity, trajectory, or an event. The exact available result types depend on the service and specification scope [6], [7].

Sensing contextual information enables interpretation of observations and results. Observation time, coordinate reference, sensing area, target assumptions, and achieved quality are especially useful for cross-node fusion. The target sensing service area is the geographic region within which the requested object or environmental characteristics are to be derived [6]. It differs from communication-cell coverage because passive-object observability depends on target scattering, geometry, clutter, receiver sensitivity, and resource availability.

| Topic and cited source | Status | Remark |
|---|---|---|
| **Service foundation**<br>**TS 22.137 V19.1.0 [6]; TR 22.837 V19.4.0 [5]** | Specified requirements and supporting study | Service objectives require radio and deployment validation. |
| **5GS architecture**<br>**TS 23.137 V20.0.0 [7]** | Stage-2 specification | gNB aerial-object framework; some SEN2 and reporting details remain unfinished. |
| **Security and privacy**<br>**TR 33.777 V20.0.0 [8]** | Completed study report | Study findings; the TR itself is not a normative security protocol. |
| **Vertical applications**<br>**TR 23.700-15 V20.1.0 [9]** | Completed study report | Application enablement; the TR itself is not a complete API specification. |
| **Channel modeling**<br>**TR 38.901 V19.5.0 [10]** | Completed study report | Target/background channel models for evaluation. |
| **NR sensing evaluation**<br>**TR 38.765 V20.0.0 [11]** | Completed study report | Monostatic UAV evaluation and processing boundaries; not a complete normative sensing protocol. |
| **6G requirements**<br>**TR 38.914 V20.1.0 [12]** | Completed study report | Scenario-specific targets beyond UAV detection and tracking. |
| **6G radio design and RF support** | Ongoing 6G study | Sensing resources, coordination, RF/coexistence and testing support remain under study. |
| **6G architecture and modes**<br>**TR 23.801-01 V0.9.0 [13]** | Ongoing 6G study | Alternative RAN/UE roles and modes; selection responsibilities and signaling remain under study. |
| **Separate SenF-C/P NFs and interface [13]** | Documented proposal | Existing 5GS internal functionalities have no standardized interface; separate NFs are not an agreed 6G requirement. |
| **Non-3GPP sensing information [13]** | Study alternatives | Collection and integration approaches; actual fusion algorithms are outside the SA2 study scope. |

**Table 1: Status of the 3GPP ISAC documents and concepts at the manuscript cutoff.**

| Use case family | Representative targets or scenario | Required sensing result | Dominant requirements |
|---|---|---|---|
| **Transportation and aerial** | UAVs, vehicles, pedestrians, railway obstacles | Detection, position, velocity, trajectory | Coverage, missed detection, false alarms, accuracy refresh rate |
| **Industrial and logistics** | AGVs, robots, machinery, restricted areas | Position, activity, safety event | Accuracy, latency, availability, continuity |
| **Human-centric** | Presence, gesture, movement, contactless monitoring | Occupancy, motion, activity class | Fine motion sensitivity, latency, privacy |
| **Environmental and public safety** | Hazard, intrusion, area monitoring | Event, map, environmental state | Wide-area coverage, reliability, long-term stability |
| **Communication assistance** | Blockage, geometry, environmental state | Beam or mobility assistance information | Freshness, confidence, low processing latency |

**Table 2: Representative 3GPP ISAC use cases and sensing requirements.**

The principal use cases can be grouped into: a) transportation and aerial sensing, b) industrial and logistics sensing, c) human-centric sensing, and d) environmental and public safety sensing, as summarized in Table 2. These applications require different output abstractions. A protected airspace service may require a three-dimensional (3D) UAV track with velocity and existence confidence. A factory safety service may require an event indicating that an object entered a restricted region. A gesture recognition service may require a semantic class rather than a globally referenced position. An environmental monitoring service may require an aggregate spatial map updated over minutes rather than a low-latency target track [5], [6].

### *B. Sensing Modes and Observability*

The sensing geometry is determined by the placement of the transmitter and receiver. In monostatic sensing, they are co-located at one RAN node or UE, and delay corresponds to the outward and return propagation between that node and the target. A common clock simplifies timing. Monostatic geometry permits either overlapping or time-separated transmission and reception. Overlapping operation requires sufficient RF isolation and, where needed, cancellation of local transmitter leakage. Time-separated operation instead introduces transmit-receive switching, receiver recovery, and listening-window constraints.

In bistatic sensing, the transmitter and receiver are at different locations. The measurement depends on the combined transmitter-to-target and target-to-receiver path. The remote receiver avoids local leakage from that sensing transmitter, although a strong direct transmitter-receiver path can mask echoes. Accurate node positions, timing and frequency alignment, configuration coordination, and knowledge of the transmitted signal are needed to interpret the observation.

Multistatic sensing combines several transmitters, receivers, or both. Complementary views can reduce geometric ambiguity and improve resilience to blockage and unfavorable target orientation. The additional observations require synchronization, calibration, transport, association, and fusion. Different nodes may observe different scattering centers on the same target, while sharing clutter or model errors. Multiple gNBs and aggregation of their sensing data are already supported by the 5GS architecture [7].

### *C. Performance Metrics and Service Requirements*

Sensing performance contains several coupled dimensions. Detection probability and false-alarm metrics depend on detector threshold and the evaluation population. A lower threshold increases sensitivity to weak targets but also increases candidate reports and downstream processing. Estimation accuracy describes error in a target parameter; resolution describes whether nearby components can be separated. Coordinate frames and observation timestamps are essential: estimates in local antenna frames require transformation before fusion, and a moving target may require prediction to the delivery time.

For conventional monostatic processing, range resolution is approximately c/(2B), where c is propagation speed and B is effective sensing bandwidth; 300 MHz gives about 0.5 m. Velocity-bin spacing scales approximately as wavelength divided by twice the coherent observation duration. A longer observation improves Doppler separation but increases sensitivity to acceleration, oscillator drift, and phase noise. Angular separation depends on wavelength, physical aperture, and calibrated spatial sampling. Array phase, antenna position and orientation, and beam-dependent RF responses affect achievable accuracy.

Service-level requirements include delivery latency, refresh period, availability, confidence, and target-area coverage. End-to-end latency includes observation, local processing, transport, fusion, result generation, and exposure. Refresh period determines how often a new result is delivered. Coverage must be qualified by target class, height, aspect, clutter, geometry, and the required detection and estimation quality. A mature service needs feasibility assessment, request rejection or renegotiation, and reporting of achieved quality when the requested performance cannot be sustained.

Service targets describe requested outcomes, including detection, accuracy, coverage, reporting period, and delivery deadline. System choices include sensing mode, participating nodes, bandwidth, observation duration, processing placement, and reporting detail. Geometry, scattering, clutter, aperture, and RF capability constrain physical feasibility; compute, transport, energy, installation, and operating cost constrain deployment viability. Standardization scope alone establishes neither physical performance nor commercial readiness.

| Metric | NR UAV study [11] | 6G Indoor Factory [12] | 6G Urban Macro [12] |
|---|---|---|---|
| **Detection probability** | 95% | 95% | 95% |
| **False-alarm probability** | 5% | 5% | 5% |
| **Horizontal localization error (90th percentile)** | 10 m | 2 m | 5 m |
| **Vertical localization error (90th percentile)** | 10 m | N/A | 8 m |
| **Velocity error (90th percentile)** | 5 m/s | 2 m/s | 4 m/s |

**Table 3: Representative sensing objectives in the 5G NR study and the 6G requirements.**

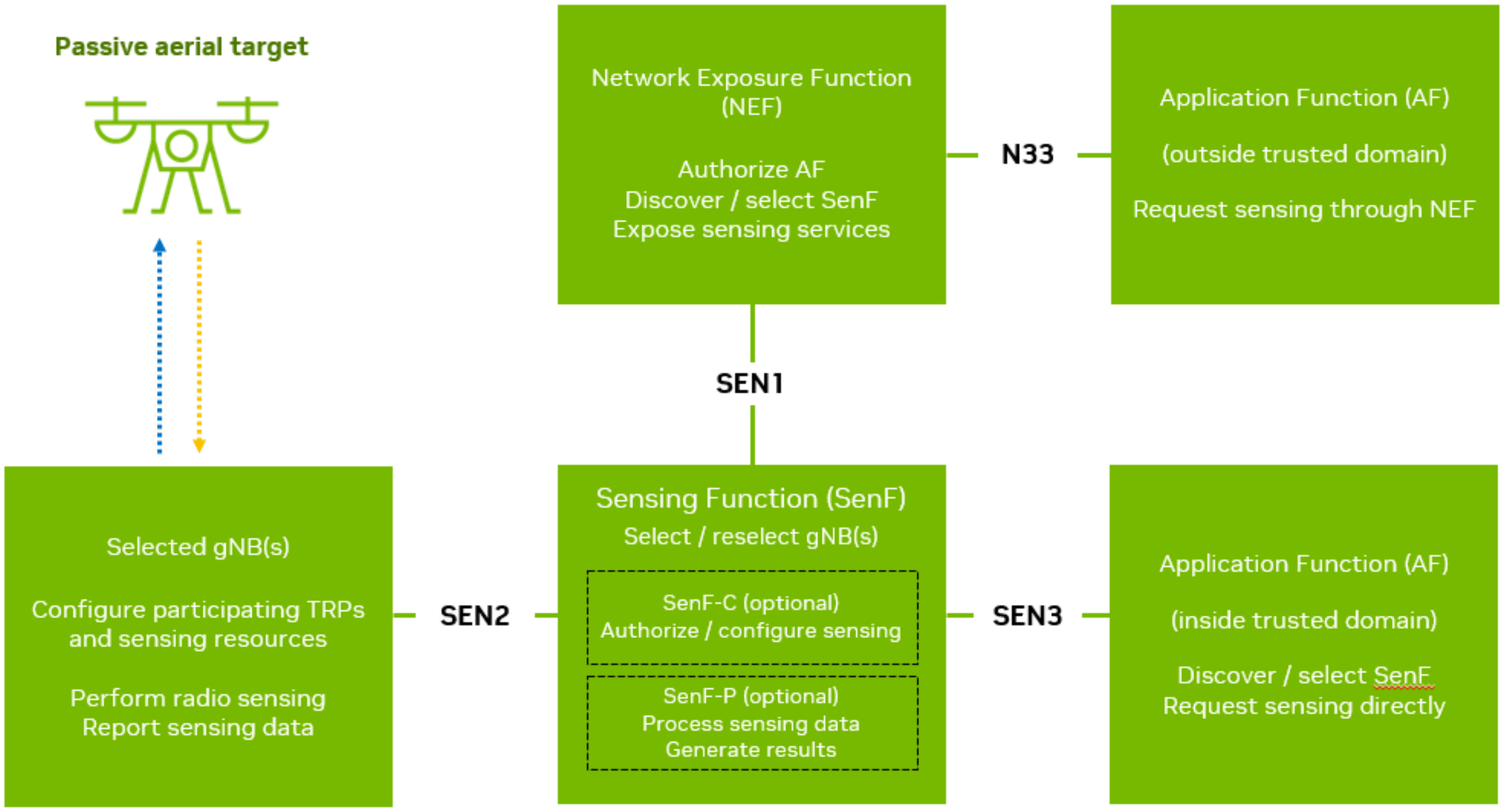

**Figure 2: 5G-Advanced Sensing Function architecture.**

### *D. From the NR Baseline to 6G Sensing Objectives*

The 5G-Advanced architecture supports aerial-object detection and tracking with gNB sensing entities [7], and the NR radio study evaluates gNB-based monostatic UAV sensing [11]. The 6G requirements broaden passive-object sensing to at least UAVs, humans, vehicles, and AGVs, and include communication assistance [12]. Environmental information may assist beam selection, blockage prediction, or mobility preparation. The 6G architecture study considers alternative solutions for RAN monostatic, RAN-RAN bistatic, UE-to-RAN bistatic, RAN-to-UE bistatic, UE monostatic, and UE-UE bistatic modes [13]. Table 3 compares the 5G NR study objectives with the 6G scenario requirements. Detection has the same numerical target, while the 6G scenarios seek tighter localization and velocity errors. The entries concern different scenarios, target populations, and evaluation assumptions.

Several improvements can be pursued within the existing framework: better clutter rejection, calibrated arrays, resource scheduling, and implementation-specific multi-gNB fusion. Other requirements motivate additional support. UE and bistatic participation need transmitter-receiver coordination, assistance information, synchronization, mobility handling, and suitable measurement reporting. Predictable short response times and fine separation of nearby objects motivate evaluation of sensing resource patterns, coherent observation support, beam procedures, and bandwidth options. These motivations identify design questions for 6G.

## III. Sensing Architecture

### *A. Sensing Function and Service Lifecycle in 5G-Advanced*

The Release-20 5G-Advanced architecture introduces a logical sensing function (SenF), as illustrated in Figure 2. TS 23.137 specifies aerial-object detection and tracking using gNB sensing entities [7]. It builds on network exposure and service-based principles. SenF responsibilities include sensing-request authorization, gNB selection and reselection, configuration, sensing-data collection and processing, result generation, fulfillment checking, and exposure. Its optional sensing-control and sensing-processing functionalities, SenF-C and SenF-P, are logical internal functionalities without a standardized interface. Placing processing near the RAN is a deployment option that can reduce transport distance and data volume.

A service begins with an application function (AF) request. A trusted AF can access the SenF directly, whereas an AF outside the trusted domain accesses it through the network exposure function (NEF). The NEF authorizes the AF, while the SenF authorizes the sensing request. The AF or NEF discovers and selects a SenF through the network repository function or local configuration. The request identifies detection or tracking, a target area, a time window, and one-time, periodic, or event-triggered reporting. It can include requested sensing performance parameters: position accuracy, velocity accuracy, and missed-detection and false-alarm probabilities. The complete mapping from service intent to radio resources remains an implementation choice. Purpose, target category, retention, and result detail are useful additional policy dimensions for deployments; their enforcement should follow the applicable security and privacy framework [8].

After authorization, the SenF selects or reselects one or more gNB sensing entities. Within a selected gNB, the gNB determines the participating transmission and reception points (TRPs) and radio resources [11]. A TRP is a radio transmission/reception point in the RAN description; its mapping to radio units and antenna panels is implementation dependent. This division explains why Figure 2 associates TRP selection with the gNB. The gNB maps the service configuration onto signals, beams, antenna resources, observation intervals, receiver processing, and the reporting split. It performs sensing and reports data or results as configured. The SenF may aggregate observations from several gNBs and generate the application-facing result.

| Abstraction | Representative information | Transport load | Processing and fusion implications |
|---|---|---|---|
| **A: Complex channel level** | Estimated complex channels in time, delay, or frequency domains | Very high | Maximum centralized flexibility; substantial calibration and privacy burden |
| **B: Parameter profile level** | Delay, Doppler, angle, or power profiles | High | Preserves multi-target structure; requires common binning and normalization |
| **C: Detected point level** | Range, velocity, angle, power, timestamp, uncertainty | Moderate | Supports centralized association and fusion; local detection may discard weak components |
| **D: Object level** | Object position, velocity, confidence; optional class and lifecycle metadata | Low | Direct application use; greatest dependence on local tracking and classification |

**Table 4: Sensing information abstractions and their transport, processing, and fusion tradeoffs.**

A sensing report includes identifiers and, where applicable, optional location, detection time, motion, and reported performance information, with event information for event-triggered reporting [7]. An ongoing service can be modified, terminated, or affected by sensing failure. The SenF can reselect gNBs as conditions change; if no suitable entity can be selected, the request is rejected or the ongoing service is notified according to the procedures [7]. A useful implementation maintains a context linking authorization, requested performance, selected gNBs, operation identifiers, reporting state, and processing state. Sharing suitable observations among authorized applications may avoid duplicate radio operations, provided their policies and performance requirements are compatible.

### *B. Processing Placement and Reporting Abstractions*

Processing placement determines how much sensing information leaves the RAN and where target interpretation occurs. TR 38.765 discusses four levels, corresponding to estimated complex channels, parameter profiles, detected points, and object-level results [11]. Table 4 summarizes their implementation tradeoffs. Table entries describe representative information and qualitative transport demands; volume depends on dimensions, quantization, and cadence.

- **Level A reporting** carries estimated complex channels in time, delay, or frequency domains. It preserves substantial flexibility for centralized processing and algorithm changes, at high transport and calibration cost.
- **Level B reporting** carries parameter profiles, such as delay, Doppler, angle, or power profiles. It can retain multi-target structure at lower volume than complex-channel reporting.
- **Level C reporting** carries detected point-related parameters such as range, velocity, azimuth, elevation, reflection power, and timestamp. Additional uncertainty information supports association and fusion. It is compact enough for network transport while preserving inputs to centralized clustering and tracking. Local detection, however, may discard weak components before cooperative evidence becomes available.
- **Level D reporting** carries object-level information produced by local clustering or tracking. It offers compact application-oriented reporting and can exploit local clutter history and calibration. Central processing receives less of the underlying measurement evidence, making correction of local association errors more difficult.

Table 4 illustrates the relationship between reporting detail and fusion flexibility. A deployment might use routine object reports, point reports for cooperative fusion, and selected profiles or channel estimates for calibration or difficult scenes. Adaptive retrieval needs capability exchange, data retention, policy control, and rules for associating supplementary evidence with the original observation.

### *C. Architectural Extensions under Study for 6G*

The 6G architecture study broadens the scope to RAN and UE sensing entities, mode selection, entity discovery and reselection, authorization, mobility handling, data transport, and result exposure [13]. Figure 3 illustrates the six sensing modes considered. UE involvement introduces mobility, energy, capability, and trust constraints beyond a fixed-gNB deployment. A UE can change serving nodes, enter power saving, or lose coverage during sensing. The architecture needs to coordinate its role as transmitter or receiver, establish applicable authorization and consent, and preserve observation context across configuration changes. Candidate mode choice must consider geometry, synchronization, radio load, target requirements, and device availability, while the responsible RAN entities retain control over feasible radio operations.

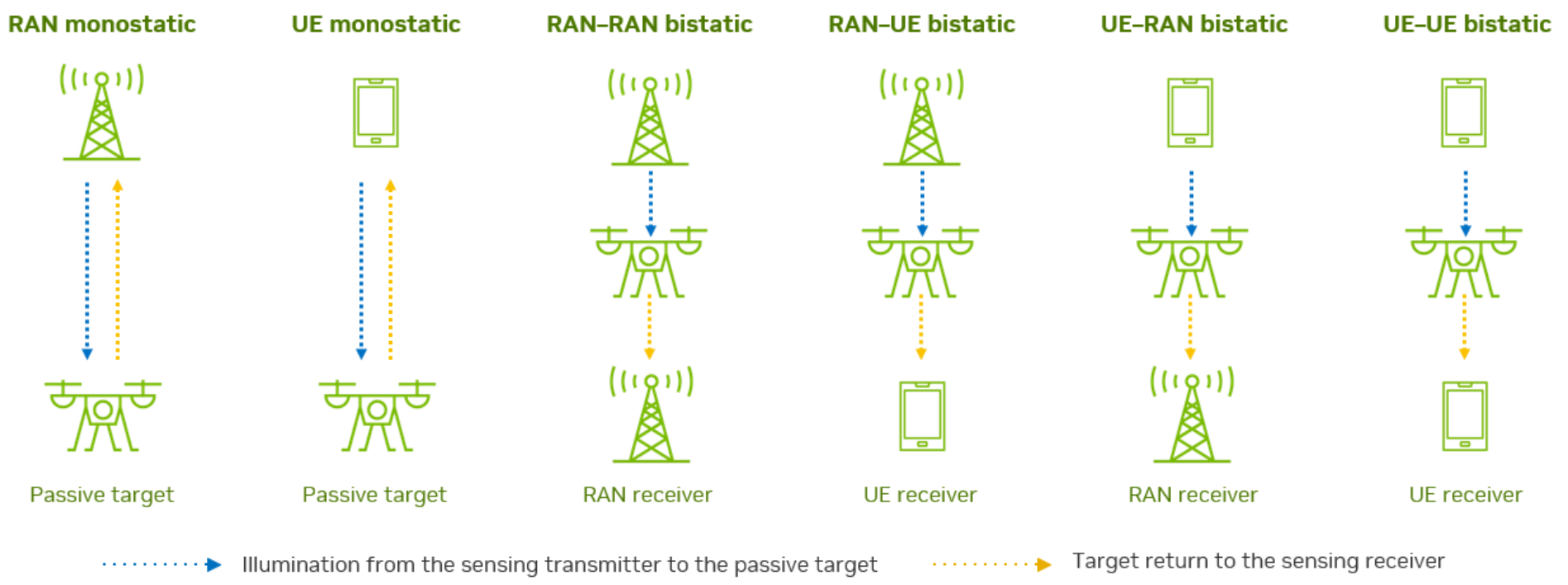

**Figure 3: Monostatic and bistatic sensing modes involving RANs and UEs; multi-static sensing combines multiple such links.**

| Area | Reusable foundation | Extension or unresolved issue |
|---|---|---|
| **Service control** | AF/NEF exposure, request authorization, lifecycle | UE roles and mobility; mode selection; broader policy context |
| **Entity management** | SenF gNB selection and reselection | RAN/UE capability discovery and responsibility across proposed solutions |
| **Data and processing** | Levels A-D abstractions; implementation-specific multi-gNB aggregation | Cross-source semantics, calibration, provenance; proposed standardized SenF-C/P interface |
| **RAN interaction** | gNB decides TRPs and resources; SEN2 reference point | Remaining SEN2 details; bistatic and UE coordination, timing and assistance |
| **Performance control** | Requested performance and fulfillment checking | Feasible mapping to radio/compute/transport; dynamic degradation and admission |

**Table 5: Reusable sensing foundations and extensions toward 6G.**

Heterogeneous sources broaden the information model, i.e., non-3GPP sensing information, including camera, lidar, Wi-Fi, and sonar observations [13]. Their timestamps, coordinate frames, calibration, and provenance must be aligned for meaningful interpretation. Sources may differ in ownership, integrity, retention, or consent guarantees. Communication assistance adds an internal consumer whose useful timescale may be shorter than that of an external surveillance application [12]. Beam management and mobility decisions require sufficiently fresh results and suitable quality indicators. Interfaces, timing budgets, and feedback into radio control remain work for the relevant studies.

### *D. Reuse, Extensions, and Incremental Deployment Cost*

A 6G sensing framework can reuse the 5GS service lifecycle, exposure mechanisms, and the distinction between sensing data and results. Extensions are needed when entities move, roles change, or information crosses technology and trust domains. Table 5 separates reusable foundations from the support still required. Reuse at the logical-function level does not imply that the associated interfaces or data semantics are already complete.

The industrial benefit is conditional reuse of sites, licensed spectrum access, downlink signals, antennas, power, backhaul, edge platforms, and operator-managed exposure. This can reduce the need to deploy a separate sensing network when existing geometry and radio hardware provide useful observations. It is especially attractive for augmenting wide-area situational awareness or filling gaps between dedicated sensors. Existing communication coverage alone does not ensure the detection reliability or spatial resolution required for a factory safety function.

Incremental costs can include sensing-capable receive chains, antenna separation or additional panels, RF isolation, array and site calibration, and compute acceleration. Sustained sensing also consumes energy, scheduling opportunities, memory, transport capacity, and operational effort. A deployment needs monitoring, RF-impairment validation, security controls, and maintenance of calibration and background models. Point or object reporting can constrain transport, but it moves processing and its lifecycle cost toward the sensing entities.

Dedicated radar, camera, or lidar can offer purpose-selected placement, sensing-specific waveforms, independent availability, and application-specific resolution. ISAC may complement these sensors or consolidate selected functions. A fair comparison therefore considers target-level coverage and reliability, capital and operating costs, spectrum and compute opportunity costs, integration effort, and the value of shared infrastructure.

## IV. RADIO SENSING

The credibility of cellular sensing depends on two tightly coupled elements: a channel model that preserves the relationship among transmitter, target, receiver, and environment, and a receiver that converts known radio transmissions into target related measurements. The ISAC channel model extensions to TR 38.901 and the 5G ISAC study in TR 38.765 provide the initial 3GPP framework for connecting target-aware propagation with radio processing [10], [11]. The Release-20 6G studies broaden this framework to additional targets, sensing modes, and RF conditions.

### *A. ISAC Channel Modeling*

A communication channel model describes the composite propagation between a transmitter and an intended receiver. For sensing, the model must retain enough physical structure to identify which components are associated with the target. A useful model therefore separates the target channel, the background channel, and local transmitter-to-receiver coupling. The target channel includes propagation to the target, the target's scattering response, and propagation to the sensing receiver. The background channel represents buildings, terrain, vegetation, machinery, vehicles, and other non-target objects. Local coupling represents direct transmitter leakage and reflections from the sensing site in monostatic sensing. TR 38.901 provides the target and background propagation framework, whereas local transmitter leakage and coupling require an additional RF or implementation impairment model.

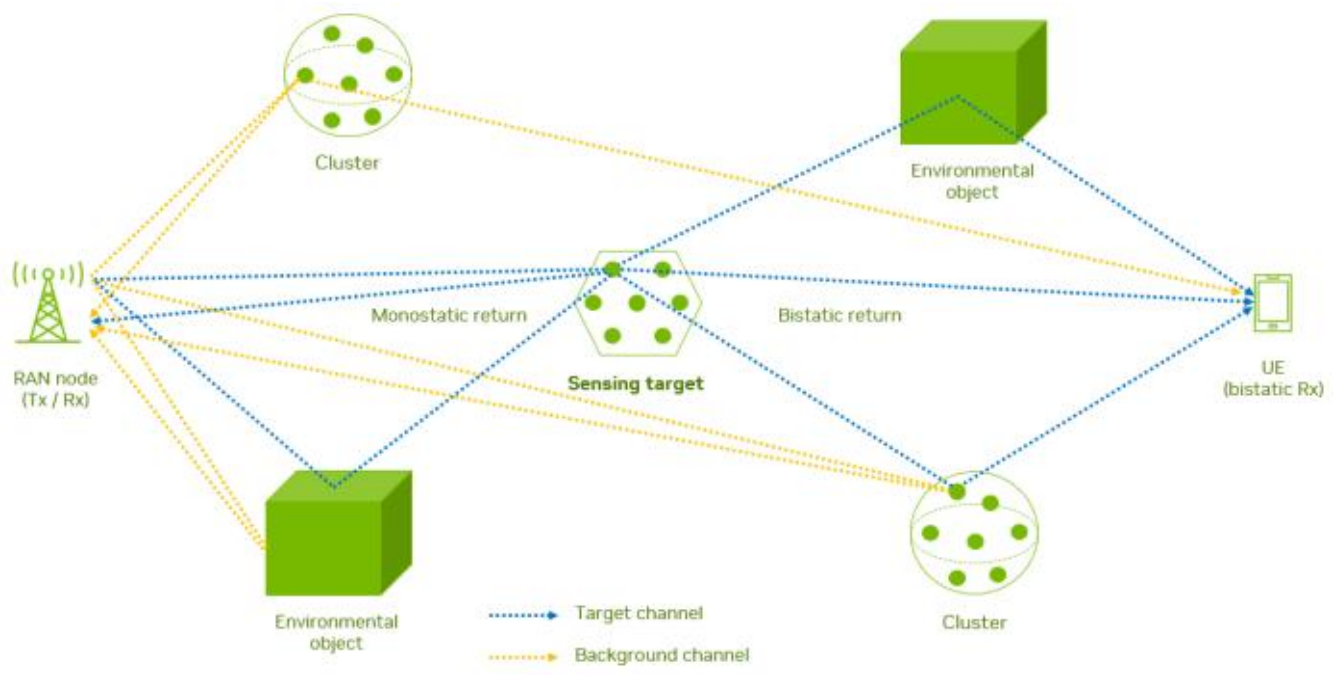

**Figure 4: An illustration of 3GPP ISAC channel model comprising target channel and background channel.**

Separating target and background components is essential for evaluating detection, as illustrated in Figure 4. The receiver distinguishes targets through delay, Doppler, angle, reflection strength, polarization, and temporal behavior, while missed detections and false alarms depend on background structure and dynamics. The geometries must remain consistent with one physical target state across time and links. Independently generating multistatic links can create unrealistic diversity and exaggerate cooperative sensing gains.

3GPP TR 38.901 represents a physical target using one or more target scattering points [10]. A scattering point is an abstraction for one or several nearby physical scattering centers with a common position, motion, RCS behavior, and polarization response. A single point may be sufficient for a compact target at long range. Extended targets require several points when they occupy multiple delay or angular resolution cells. Vehicles, humans, AGVs, and UAVs therefore require different levels of target detail. Besides, RCS should not be treated as one fixed value for a target class. It varies with frequency, material, shape, polarization, target orientation, and illumination and observation angles. A UAV can move through deep reflection minima even when its range changes only slightly. This aspect dependence is important for tracking because a valid target may disappear temporarily from one view and reappear as its orientation changes. The model should preserve consistency between target motion, orientation, and reflection strength over time.

Extended and articulated targets may also produce micro-Doppler. Detailed micro-Doppler is particularly relevant to classification studies and may require target models beyond the baseline detection-oriented representation. Rotating UAV propellers, vehicle wheels, human limbs, and industrial machinery create Doppler components around the bulk target motion. The required detail depends on the study objective. Basic detection may need only a dominant target component and realistic power variation. Tracking needs temporally consistent reflection changes. Classification requires richer representation of moving target parts. The background channel requires equal care. Static clutter includes buildings, terrain, parked vehicles, and fixed infrastructure; dynamic clutter includes traffic, people, birds, and rotating equipment. These components should evolve consistently across time, beams, and sensing links. Such consistency is necessary for evaluating receivers that learn or subtract a long-term background [14].

### *B. Radio Processing*

For the gNB-based monostatic baseline studied in TR 38.765, the receiver operation can be organized into the following five tutorial steps [11]. This organization explains implementation dependencies; the study does not prescribe one mandatory five-stage algorithm.

**Step 1: transmitted-signal reconstruction:** The first requirement is accurate reconstruction of what was radiated. A gNB knows the downlink reference sequences and transmitted data symbols, but the sensing processor must also account for resource allocation, physical channel mapping, precoding, and beamforming. This is particularly important when data-bearing resources are reused. They can increase effective bandwidth and energy, but communication scheduling creates gaps and may change the precoder or beam across time and frequency. Those changes alter target illumination and must not be interpreted as target motion.

**Step 2: calibration and leakage management:** Calibration removes deterministic radio and array effects. Frequency-dependent responses can distort delay profiles, timing and frequency offsets can appear as range and velocity biases, and antenna phase errors can bias angle estimates. Communication-grade calibration may be insufficient because errors with little throughput impact can still bias sensing. When transmission and reception overlap, antenna separation, directional isolation, shielding, filtering, and, where needed, analog cancellation protect the receiver from strong local leakage. Digital cancellation can then reduce residual leakage using the reconstructed waveform and an estimated coupling channel, accounting for phase noise, IQ imbalance, and nonlinearities. It cannot undo front-end saturation or analog-digital-conversion (ADC) clipping. With time-separated operation, switching and recovery transients, listening windows, and unavailable echo intervals replace continuous leakage cancellation as key constraints.

**Step 3: delay-Doppler-angle processing:** After calibration and leakage suppression, the receiver estimates delay, Doppler, and angle. Frequency-domain phase variation provides delay information, temporal phase evolution provides Doppler, and spatial variation across antennas provides direction. The resulting delay-Doppler-angle response is an intermediate radio representation rather than an object list. Strong cells may correspond to the target, environmental clutter, sidelobes, or residual leakage. The actual 5G resource pattern matters. Sparse subcarriers create delay sidelobes and ambiguities; gaps between sensing occasions degrade Doppler processing; and beam or precoder changes couple spatial and temporal observations. Windowing can suppress sidelobes from strong components, but it broadens the main response and makes nearby targets harder to separate.

**Step 4: clutter suppression and detection:** Clutter suppression is then applied to the delay-Doppler-angle response. Static background subtraction can reduce persistent environmental components, while temporal or Doppler filtering emphasizes moving targets. A practical receiver may combine a moving-target branch with a slower scene-change branch that compares the current observation with a long-term environmental map. Detection identifies statistically significant components in the clutter-suppressed response. Adaptive thresholds, including constant-false-alarm-rate (CFAR) techniques, are useful because the background varies across range, direction, and environment. The threshold controls the balance between weak target sensitivity and false alarms.

**Step 5: target-parameter estimation:** The radio processor estimates parameters for each detected component. A point-level measurement may include delay or range, Doppler or radial velocity, azimuth, elevation, reflection strength, and timestamp, together with uncertainty or quality metadata. Extended targets may produce several points. Preserving these measurements allows later clustering, tracking, classification, and multi-node fusion rather than forcing a radio detector to declare one object. Their implementation requires consistent coordinate and time references, calibration, quantization, capability exchange, and reporting semantics; object-level reporting is also possible when interpretation is placed locally.

The 6G radio work can revisit these dependencies. Suitable sensing resources may improve delay-Doppler ambiguity and coherent integration. Bistatic sensing avoids the sensing transmitter's local leakage at the remote receiver but requires synchronization, signal knowledge, and direct-path management. Cooperation needs consistent measurements and association across nodes.

### C. Bandwidth, Frequency, and Radio Design Tradeoffs

Wider effective bandwidth can separate nearby path components and may help distinguish targets from clutter in factories or dense traffic. Its value depends on the occupied frequency pattern: a large fragmented span can introduce ambiguities and sidelobes, while a contiguous span provides a different response. Higher sampling rates increase memory traffic, transport if early-stage data are exported, and processing cost. Accurate timing and frequency-dependent calibration become more demanding. Improved range resolution alone does not ensure better 3D localization when angle or geometry dominates the error.

Higher carrier frequency reduces wavelength. For a fixed physical aperture, this permits a narrower angular response and potentially more spatial samples. With a fixed number of half-wavelength-spaced elements, however, aperture also shrinks, so frequency alone does not guarantee finer angular resolution. Narrow beams can support localized sensing and communication assistance, while increasing beam-search and tracking overhead. Blockage, propagation loss, aspect-dependent scattering and RCS, phase noise, RF dynamic range, and calibration can offset the nominal benefit. Extended targets may also require richer frequency- and aspect-dependent channel models. The terrestrial frequency range in the 6G radio study is 410 MHz–52.6 GHz [12]. Sub-terahertz sensing is a broader research direction beyond that frequency scope.

Fine velocity separation favors longer coherent observations, whereas short response times and rapid target dynamics favor shorter intervals. Waveform and resource design must balance these needs with communication scheduling and available receive windows. Enhanced signals or resource patterns merit evaluation when existing configurations cannot meet a specified service budget under realistic impairments.

## V. END-TO-END CASE STUDY OF COOPERATIVE UAV MONITORING

### A. Service Objectives and Configuration Choices

Consider an operator-managed service that reports UAV positions in an authorized area. For illustration, the application requests detection probability of at least 95%, 90th-percentile 3D position error no greater than 5 m, a 200 ms reporting period (5 Hz), and delivery within 200 ms of the start of each observation window. The deadline applies to an established periodic operation; initial authorization, entity setup, and waiting for the first observation window are additional delays.

The service first requires feasible coverage and complementary observations. The SenF selects candidate gNBs for the target area, and each selected gNB determines its TRPs and resources. We consider cooperative monostatic sensing at three sites. We place channel estimation, clutter rejection, and candidate detection near each gNB, then associate and fuse point reports at an edge processing instance serving the selected area. It exploits local signal knowledge while limiting the transport of complex channels. The AF receives object-level position and motion information through the authorized exposure path. The three choices (i.e., local measurement processing, cross-site point fusion, and application-level results) connect the reporting abstractions of Section III to distinct transport and processing needs. Table 6 illustrates the mapping from UAV service objectives to implementation decisions.

| Service objective | Configuration and feasibility check |
|---|---|
| **Detection and coverage** | Select complementary gNB/TRP views; retain useful weak candidates; verify detection over the target area. |
| **5 m p90 3D error** | Calibrate geometry and timing; fuse point evidence with uncertainty; assess the population on which accuracy is reported. |
| **200 ms deadline** | Budget observation, local processing, transport, fusion, and exposure; reserve time for any retrieval round trip. |
| **5 Hz reporting** | Schedule repeated windows and processing capacity; timestamp results and qualify their age. Continuous tracking needs separate validation. |
| **Transport and privacy** | Keep high-rate intermediate data local; send selected points for fusion and appropriate object-level results to the AF. |

**Table 6: Illustrative mapping from UAV service objectives to implementation decisions.**

### B. Evaluation Setup and Results

The companion preprint [15] provides the algorithms and simulation details. The numerical setup uses 4 GHz carrier frequency, 30 kHz subcarrier spacing, and a 273-resource-block carrier. The full-band single-site configuration occupies 272 resource blocks. The three-site configuration uses 92/92/88 resource blocks at 47.229 dBm per site, versus 52 dBm at the full-band site. Aggregate transmitted reference-signal resources and energy are matched.

Each nominal 128 ms coherent processing interval (CPI) contains 32 sensing occasions spaced 4 ms apart, with occasion starts from 0 to 124 ms. The mapped evaluation contains 1,024 occupied scenes with five UAVs each (5,120 target instances) and 1,024 empty scenes. Detection scoring uses one-to-one association within 30 m in 3D, first maximizing matched targets and then minimizing total error. Table 7 separates local detection checks, cooperative outcomes, and conditional positioning accuracy.

Local validation checks reduce false candidates at the cost of missed targets. The reference receiver already uses clutter subtraction, windowing, thresholding, and peak separation. In the cooperative result, Type 2 is an average of per-scene false-output fractions [11]. Observing no empty-scene alarm does not establish zero underlying probability: 0/1,024 gives a one-sided 95% binomial upper bound of approximately 0.292%.

The positioning comparison uses the fixed full-band site selected after evaluation for the highest overall detection rate. It is a different baseline from the reference site in the local-check comparison. The 3.50 m p90 error applies only to the 4,511 targets detected by both systems. Cooperative detection of 92.46% falls below the illustrative 95% objective, so the results do not establish joint service compliance.

| Comparison or metric | Reported outcome and population |
|---|---|
| **Local validation checks, reference site** | Detection: 84.20% → 80.59%; mean unmatched outputs: 8.742 → 1.904 per occupied scene. |
| **Cooperative mapped detection** | 4,734/5,120 targets, or 92.46%; 22 unmatched outputs across occupied scenes. |
| **Cooperative false alarms** | NR-style Type 2: 0.396% per-scene average; empty-scene alarms: 0/1,024. |
| **Position error on common detections** | p90 3D error: 8.54 m → 3.50 m, best fixed full-band site versus cooperation, on 4,511 commonly detected targets. |
| **Selective weak-point reporting** | Weak uploads: 1,558 → 213, with 2,966 receiver-directed requests; 13,434 strong uploads are common to both policies. |

**Table 7: Selected results from the companion preprint [15].**

### C. *Reporting Overhead and Latency*

Early channel or profile reporting preserves weak evidence for centralized processing but consumes transport and shifts calibration and compute requirements toward the fusion platform. Point reporting reduces volume while allowing cross-site association. Aggressive local filtering can remove measurements that another site could corroborate. Object-level reporting is more compact, but local association or tracking decisions may hide useful alternatives. The preprint [15] examines a selective policy that retains weak candidates locally and uploads them when requested by the fusion process.

The weak-upload count falls by 86.3%, from 1,558 to 213, but the policy also generates 2,966 requests. Let M be the bytes per point record and Q the bytes per request. Including 13,434 strong uploads common to both policies, proactive reporting requires 14,992M bytes and selective reporting requires 13,647M + 2,966Q bytes. Selective reporting reduces this modeled total only when $Q/M < 1{,}345/2{,}966 \approx 0.453$. Equal-sized requests and point records increase total traffic by 10.81%; requests one quarter the size of a point record reduce it by 4.03%.

The architectural implication is that reporting abstraction and retrieval policy must be evaluated together. A request may be small, batched, or carried on an existing control exchange, but it consumes processing and transport time. Selective retrieval is attractive only when saved data and its sensing value justify the request cost and the additional round trip. Proactive point reporting may be preferable when the deadline leaves little room for feedback.

For the illustrative 200 ms start-of-window-to-delivery deadline, a nominal 128 ms observation leaves at most 72 ms for the remaining serial stages: local processing, transport, any request/response exchange, fusion, result generation, and exposure. Some stages can overlap in an implementation, but their execution and queueing delays must be measured before making a commitment.

At 4 GHz, the ideal monostatic velocity-bin spacing is approximately 0.29 m/s for 128 ms and 0.59 m/s for 64 ms. Halving the observation duration creates more time for downstream work but reduces coherent evidence and Doppler separation. A practical controller would assess the configured observation and processing pipeline against both the update period and deadline, including load-dependent delay. If the service is infeasible, choices include more suitable entities, additional resources, a different reporting policy, or a revised objective. The simulated sensing quality, the analytical traffic model, and the proposed latency budget address different parts of this decision.

## VI. CONCLUSIONS AND FUTURE OUTLOOK

3GPP ISAC is evolving from an exploratory radio capability into an end-to-end network service. Release 19 established use-case, service, and channel-model foundations. Release 20 develops the 5G-Advanced aerial-object sensing architecture and NR monostatic evaluation baseline, while 6G studies broaden targets, sensing roles, modes, and information sources. The 6G requirements target tighter scenario-specific localization and velocity errors. A practical evolution can reuse service control and exposure while extending coordination, measurement semantics, and radio support where justified by application needs and implementation evidence. The UAV case study shows how those needs affect entity selection, processing placement, and reporting abstraction.

We conclude by pointing out some fruitful avenues that require further work.

**Sensing-communication co-design:** 6G sensing must progress beyond opportunistic reuse of communication signals. Sensing-aware scheduling, beam management, and admission control are needed to provide predictable sensing quality without excessive communication overhead. Communication assistance results must also be delivered within the timescale of beam, mobility, and scheduling decisions.

**Cooperative and multimodal sensing:** Multi-node and multimodal sensing can improve coverage, accuracy, and resilience by combining radio, camera, lidar, or other observations. These gains require aligned timestamps, coordinate frames, calibration states, and provenance, together with robust cross-node and cross-modal association.

**Security and privacy:** Passive targets may include individuals with no subscription or signaling relationship with the network. Authorization should therefore constrain the sensing area, target category, result detail, purpose, duration, and retention policy. Edge processing and minimum necessary result exposure can reduce the risks associated with transporting detailed radio observations.

**Reproducible evaluation:** Fair comparison requires shared target and clutter models, RF-impairment assumptions, and ground truth methods. These elements should be complemented by open-source ISAC evaluation tools that implement reference waveforms, channel and target models, receivers, baseline algorithms, and reference outputs. They should support simulation, measured datasets, and hardware-in-the-loop validation, enabling repeatable results.